\documentclass[12pt]{article}
\usepackage{epsfig}

\def \D {\hbox{d}}

\begin{document}

\title{
A connection between HH3 and KdV with one source\footnote{
Corresponding author Zhao, fax +86-10-88256100 Preprint S2009/???.}}

\author{Jun-xiao Zhao$^{1,3}$
 and Robert Conte$^{2,3}$
\\
\\
$^{1}$School of Mathematical Sciences, Graduate University of
\\ Chinese Academy of Sciences, Bejing, 100049, P.R.~China
\\
\\
$^{2}$LRC MESO, ENS Cachan et CEA/DAM 
\\ 61, avenue du Pr\'esident Wilson,
  F--94235 Cachan Cedex, France.
\\
\\
$^{3}$Service de physique de l'\'etat condens\'e (URA 2464)
\\ CEA--Saclay, F--91191 Gif-sur-Yvette Cedex, France
\smallskip
\\
\\ E-mail:  Robert.Conte@cea.fr
\\
E-mail:  jxzhao@gucas.ac.cn } {} 
\maketitle

\begin{abstract}

In the system made of Korteweg-de Vries with one source, 
we first show by applying the Painlev\'e test that the two components
of the source must have the same potential. 
We then explain the natural introduction of an additional term
in the potential of the source equations
while preserving the existence of a Lax pair.
This allows us to prove the identity between the travelling wave
reduction and one of the three integrable cases of the cubic
H\'enon-Heiles Hamiltonian system.
\end{abstract}

\noindent \textit{Keywords}:
KdV with source, 
cubic H\'enon-Heiles,
Lax pair, 
Painlev\'e test. 
\medskip

\noindent \textit{AMS MSC 2000}
 34A05, 
 35A20. 

\noindent \textit{PACS 2001} 
  02.30.Jr,
  02.30.Ik.


\baselineskip=12truept



\section{Introduction}

In several soliton equations, it is possible to add a so-called
``source term'' without destroying the soliton property. 
For instance, the Korteweg-de Vries (KdV) equation
\begin{eqnarray}
\label{eqKdV}
& & b u_t +\left(u_{xx} -\frac{3}{a} u^2\right)_x=0,\
 a, b=\mbox{constants},
\end{eqnarray}
retains its solitonic property when extended to
\cite{Melnikov1983,ZK1986}
\begin{eqnarray}
& & \left\lbrace
\begin{array}{ll}
\displaystyle{
 E_1 \equiv b u_t +\left(u_{xx} -\frac{3}{a} u^2 - d v w\right)_x=0,\ ~
  a, b, d=\mbox{constants},
}\\ \displaystyle{
 E_2 \equiv v_{xx}+\left(-\frac{u}{a} + \mu_1 \right) v=0, }
\\ \displaystyle{
 E_3 \equiv w_{xx}+\left(-\frac{u}{a} + \mu_2 \right) w=0, 
}
\end{array}
\right. 
\label{eqKdVonesource}
\end{eqnarray}
provided $\mu_1=\mu_2=\mu$.
The parameter $\mu$ is then inessential and it could be removed by 
the Galilean transformation
\begin{eqnarray}
& &
(u,x,t) \to \left(u+ a \mu, \ x+ 6 \frac{\mu}{b} t,\ t\right). 
\end{eqnarray}

If one denotes $W=v w$, $Z=v/w$, the transformed system for $(u,W,Z)$, 
\begin{eqnarray}
& & \left\lbrace
\begin{array}{ll}
\displaystyle{
  b u_t +\left(u_{xx}  -\frac{3}{a} u^2   - d W\right)_x=0,\ 
}\\ \displaystyle{
\left(\partial_x^3 - 4 \left(\frac{u}{a}- \mu\right)\partial_x
      -2 \frac{u_x}{a}\right)W=0,
}\\ \displaystyle{
\left(W (\log Z)_x\right)_x=0,
}
\end{array}
\right.
\label{eqKdVuWZnoG}
\end{eqnarray}
allows the easy elimination of $W$ and $Z$,
resulting in the KdV6 equation \cite{KKSST,Ku2008},
\begin{eqnarray}
& &
\left(\partial_x^3 - \frac{4}{a}  U_x\partial_x- \frac{2}{a} U_{xx}\right)
\left(b U_t +U_{xxx} -\frac{3}{a} U_x^2\right)=0,\
U_x=u-\alpha \mu.
\label{eqKdV6U}
\end{eqnarray}

Solutions of system (\ref{eqKdVonesource}) with $\mu_1=\mu_2$ have
been investigated by means of the inverse scattering method
\cite{Melnikov1988} and the Darboux transformation
\cite{ZengMaShao2001,LinZengMa2001}. 
In the more general situation
$\mu_1\not=\mu_2$, it has been noticed \cite{Ma2005} that one can
also build a variety of solutions.

The purpose of this paper is to present some new results concerning
the KdV with one source system (\ref{eqKdVonesource}).
In section \ref{sectionPT}, 
we first examine the system (\ref{eqKdVonesource})
and prove that a necessary condition to pass the Painlev\'e test is
$\mu_1=\mu_2$.
In section \ref{linkHH}, 
we introduce an extension of this system admitting a Lax pair,
and
we show that this generalized KdV with one source
admits a reduction which can be identified to one of the three
integrable sets of equations of motion of the cubic H\'enon-Heiles
Hamiltonian system.

\section{The Painlev\'e test}
\label{sectionPT}

Among the possible leading behaviours
\begin{eqnarray}
& & u \sim u_0 \chi^{p_1},\ v \sim v_0 \chi^{p_2},\ w \sim w_0
\chi^{p_3},\ u_0 v_0 w_0 \not=0,
\end{eqnarray}
in which $\chi$ denotes the expansion variable near a movable
singularity $\varphi=\varphi(x,t)$ \cite{Conte1989}
\begin{eqnarray}
\chi=\left(
\frac{\varphi_x}{\varphi}-\frac{\varphi_{xx}}{2\varphi_x}\right)^{-1},
\end{eqnarray}
there exist at least two families in which all $p_i$ are negative
integers, these are
\begin{eqnarray}
& & \left\lbrace
\begin{array}{ll}
\displaystyle{ 
\hbox{F1}:\ {\bf p}=(-2,-1,-1),\ u_0=2 a,\
(v_0,w_0)=\hbox{arbitrary}, }
\\ \displaystyle{ 
\hbox{F2}:\ {\bf p}=(-2,-2,-2),\ u_0=6 a,\ v_0
w_0=-\frac{72 a}{d}, }
\end{array}
\right. 
\label{eqKdVonesourceFamilies}
\end{eqnarray}
with ${\bf p}=(p_1,p_2,p_3)$, and their respective Fuchs indices are
\begin{eqnarray}
& & \left\lbrace
\begin{array}{ll}
\displaystyle{ 
\hbox{F1}:\ -1,0,0,3,3,4,6, }
\\ \displaystyle{ 
\hbox{F2}:\ -3,-1,0,4,5,6,8. }
\end{array}
\right. 
\label{eqKdVonesourceIndices}
\end{eqnarray}
When one checks the existence of the Laurent series, one finds the
following conditions for the absence of movable logarithms,
\begin{eqnarray}
& & \hbox{F1}: Q_6 \equiv (\mu_1-\mu_2)^2 v_0 w_0 =0,
\end{eqnarray}
and
\begin{eqnarray}
& & {\hskip -10.0 truemm}
\hbox{F2}: \left\lbrace
\begin{array}{ll}
\displaystyle{ 
Q_6 \equiv (\mu_1-\mu_2) (b C+3(\mu_1+\mu_2))=0, }
\\ \displaystyle{ 
Q_8 \equiv 2 S Q_6 + (\mu_1-\mu_2) (20 S_{xx} - 20 S^2 -4 b C_{xx}
 + 200 u_4 + 5 (\mu_1-\mu_2)^2)=0, 
}
\end{array}
\right.
\end{eqnarray}
where $C=C(x,t),S=S(x,t)$ are functions given by the singular
manifold \cite{Conte1989}
\begin{equation}
\label{sc}
 S=\left(\frac{\varphi_{xx}}{\varphi_x}\right)_x
 -\frac{1}{2} \left(\frac{\varphi_{xx}}{\varphi_x}\right)^2, 
 \quad
 C=-\frac{\varphi_t}{\varphi_x},
\end{equation}
and $u_4$ is the arbitrary coefficient introduced at Fuchs index $i=4$.
Therefore a necessary condition for the system
(\ref{eqKdVonesource}) to pass the Painlev\'e test is that
$\mu_1=\mu_2$.

\section{Link with the cubic H\'enon-Heiles}
\label{linkHH}

The last two equations of system (\ref{eqKdVuWZnoG})
each admit a first integral related to the Wronskian of $v$ and $w$,
\begin{eqnarray}
& & 
2 G(t)=W W_{xx} -\frac{1}{2} {W_x}^2 - 2 \left(\frac{u}{a} - \mu\right) W^2,
\label{eqFirstIntegralG}
\\
& &
W (\log Z)_x=g(t),\ 
\label{eqFirstIntegralg}
\\
& &
g(t)=v_x w - v w_x,\ G(t)=-\frac{1}{2} g^2(t).
\end{eqnarray}
The conservation of the Wronskian is of course intrinsic to the original
system (\ref{eqKdVonesource}) (with $\mu_1=\mu_2$),
and it manifests itself by the natural introduction in the transformed
system for $(u,W,Z)$ of one arbitrary function of time,
thus ensuring the conservation of the total differential order (seven)
between the original and the transformed systems.

One deduces that the field
\begin{eqnarray}
& & 
Q=W^{1/2}
\end{eqnarray}
obeys a nonlinear ODE of the Ermakov-Pinney type \cite{Ermakov,Pinney}
\begin{eqnarray}
& & 
\left(\partial_x^2 -\frac{u}{a} + \mu - \frac{G(t)}{Q^4} \right) Q=0,
\label{eqErmakov}
\end{eqnarray}
i.e.~an equation which only differs from the linear ODE for $v$ or $w$ 
in system (\ref{eqKdVonesource}) by the contribution of $G(t)$.

The main point is that the KdV with one source system (\ref{eqKdVonesource})
is incomplete,
in the sense that the term $G(t)/W^2$ can be added
to the potential of the linear equation for $v$ and $w$
while retaining the Painlev\'e property.
Indeed, the extrapolation of (\ref{eqKdVonesource}) to an arbitrary value 
of $G(t)$ 
\begin{eqnarray}
& & \left\lbrace
\begin{array}{ll}
\displaystyle{ 
F_1 \equiv b u_t +\left(u_{xx} -\frac{3}{a} u^2 - d v w\right)_x=0,
}\\ \displaystyle{
F_2 \equiv v_{xx}+\left(-\frac{u}{a} + \mu -\frac{G(t)}{v^2 w^2}\right) v=0,
}\\ \displaystyle{ 
F_3 \equiv w_{xx}+\left(-\frac{u}{a} + \mu -\frac{G(t)}{v^2 w^2}\right) w=0, 
}\end{array} \right.
\label{eqKdVonesourceg}
\end{eqnarray}
admits the second order matrix Lax pair
\begin{eqnarray}
& & \Psi_x= L \Psi,\
 \Psi_t= M \Psi,
 \Psi=\pmatrix{\psi_1 \cr \psi_2 \cr},\
\label{eqLaxKdVSourceComplete}
\\
& &
L=\pmatrix{0 & 1 \cr \displaystyle{\frac{u}{a}}- \mu + \lambda & 0 \cr},\ 
b M=\left(-\frac{u_x}{a}-\frac{d (v w)_x}{4a \lambda}\right)
  \pmatrix{1 & 0 \cr 0 & -1 \cr}
 +\pmatrix{0 & a_{12} \cr a_{21} & 0 \cr},
\nonumber\\ & & 
a_{12}= \frac{2u}{a}- 4 (\lambda- \mu) + \frac{d v w}{2 a \lambda},
\nonumber\\ & & 
a_{21}= \left( \frac{u}{a}+  \lambda-\mu \right)
        \left(2\frac{u}{a}-4(\lambda-\mu)\right)
       -\frac{u_{xx}}{a}   
       -\frac{d G(t)}{2 a \lambda v w}
       +\frac{d v w}{2 a}
       -\frac{d v_x w_x}{2 a \lambda},
\nonumber
\end{eqnarray}
in which $\lambda$ is the spectral parameter.
The zero-curvature vanishing condition 
\begin{eqnarray}
& & {\hskip -10.0truemm}
b(L_t-M_x+[L,M])
 \equiv
 \frac {d( w F_2 + v F_3 )}{4 a \lambda} 
 \pmatrix{1 & 0 \cr 0 & -1 \cr} 
 + \pmatrix{0 & 0 \cr 
\displaystyle{ \frac{F_1}{a}+\frac{d(w_x F_2 + v_x F_3)}{2 a \lambda} }& 0 \cr}, 
\end{eqnarray}
is indeed equivalent to the condition that the lhs $F_1, F_2, F_3$ of the system
(\ref{eqKdVonesourceg}) simultaneously vanish.

Let us now prove that the system (\ref{eqKdVonesourceg}) admits for
$G(t)$ arbitrary a noncharacteristic 
reduction to an ODE system which is complete in the Painlev\'e sense.
The transformed system of (\ref{eqKdVonesourceg}) for $(u,W=v w,Z=v/w)$ 
reads
\begin{eqnarray}
& & \left\lbrace
\begin{array}{ll}
\displaystyle{
  b u_t +\left(u_{xx}  -\frac{3}{a} u^2   - d W\right)_x=0,\ 
}\\ \displaystyle{
\left(\partial_x^3 - 4 \left(\frac{u}{a}- \mu-\frac{G(t)}{W^2}\right)\partial_x
      -2 \left(\frac{u}{a}-\frac{G(t)}{W^2}\right)_x \right)W=0,
}\\ \displaystyle{
W (\log Z)_x=g(t),\ 
}
\end{array}
\right.
\label{eqKdVuWZwithG}
\end{eqnarray}
the second equation is independent of $G(t)$ 
and therefore admits the first integral $G(t)$ defined in 
(\ref{eqFirstIntegralG}),
and the first two equations define a closed susbsystem for $u,W$.
The resulting system for $(u,Q=W^{1/2})$ is
\begin{eqnarray}
& & \left\lbrace
\begin{array}{ll}
\displaystyle{
b u_t +\left(u_{xx}  -\frac{3}{a} u^2   - d Q^2\right)_x=0,\ 
}\\ \displaystyle{
\left(\partial_x^2 -\frac{u}{a} + \mu - \frac{G(t)}{Q^4} \right) Q=0.
}
\end{array}
\right.
\label{eqKdVuQwithG}
\end{eqnarray}

As to the cubic H\'enon-Heiles Hamiltonian, it is defined as
\cite{HenonHeiles,Fordy1991}
\begin{eqnarray}
 & & H=\frac{1}{2} (p_1^2+p_2^2+c_1 q_1^2+c_2 q_2^2)+\alpha q_1 q_2^2
      -\frac{1}{3} \beta q_1^3+\frac{c_3}{2 q_2^2},\ \alpha\not=0, 
\label{HHhamilton}
\end{eqnarray}
in which $p_i=p_i(\xi),~q_i=q_i(\xi)~(i=1,2)$ and
$\alpha,\beta,c_1,c_2,c_3$ are constants. 
The corresponding equations of motion,
\begin{eqnarray}
\label{HH}
\left\{\begin{array}{l}
\displaystyle{
 \frac{\D^2 q_1}{\D \xi^2}+c_1q_1-\beta q_1^2+\alpha q_2^2=0,
}\\ \displaystyle{
 \frac{\D^2 q_2}{\D \xi^2}+c_2q_2+2\alpha q_1 q_2-\frac{c_3}{q_2^3}=0,
}
\end{array}
\right.
\end{eqnarray}
pass the Painlev\'e test (in the dependent variables $q_1,q_2^2$) 
for only three sets of values of $(\beta/\alpha,c_1,c_2)$ 
\cite{BSV1982,CTW,GDP1982b}: 
$(\beta/ \alpha=-1, c_1=c_2)$, 
$(\beta/ \alpha=-6)$,
$(\beta/ \alpha=-16, c_1=16 c_2)$.
These three cases are equivalent to the stationary reduction
of three fifth order soliton equations, respectively known as the
Sawada-Kotera (SK), fifth order Korteweg-de-Vries (KdV${}_5$) and
Kaup-Kupershmidt (KK) equations, which belong respectively to the
BKP, KP and CKP hierarchies.

The link between the extended KdV with one source system 
(\ref{eqKdVonesourceg}) now becomes obvious.
Under the traveling wave reduction
\begin{eqnarray}
& & G(t)=\hbox{constant},\quad \xi=x-ct, 
\end{eqnarray}
the two-component system 
is readily identified to the Hamilton equations (\ref{HH}), with
\begin{eqnarray}
& & u=q_1,\ Q=W^{1/2}=(v w)^{1/2}=q_2,\ 
\nonumber \\ & & 
c_1= - b c,\ \beta=\frac{3}{a},\ \alpha=-d,\
c_2=\mu,\ \alpha=-\frac{1}{2a},\ c_3=G(t),\
\end{eqnarray}
which imply
\begin{eqnarray}
\frac {\beta}{\alpha}=-6.
\end{eqnarray}
Therefore the traveling-wave reduction of the extended KdV with one
source system (\ref{eqKdVonesourceg}) is the cubic HH system
corresponding to KdV${}_5$.

Since one cannot include additional terms in the system (\ref{HH})
without destroying its Painlev\'e property
\cite{Drach1919KdV,Fordy1991,CosPole2} (completeness property), 
this proves that the initial system (\ref{eqKdVonesource}) 
(with $\mu_1=\mu_2$) was incomplete.
This is why the extended KdV with one source system (\ref{eqKdVonesourceg}) 
deserves to be qualified as ``complete''.

\textit{Remark}.
As proven in Ref.~\cite{Eilbeck},
the number of degrees of freedom in (\ref{HHhamilton})
can be extended arbitrarily.

\section{Conclusion}

It is known that the cubic and quartic H\'enon-Heiles Hamiltonians
pass the Painlev\'e test only for seven sets of coefficients. 
The seven H\'enon-Heiles Hamiltonians all have the Painlev\'e property
and have been extensively studied
\cite{Blaszak1993,CMVCalogero,CMVGallipoli2004,BEF1995b}. 
However, the explicit integration of three of the quartic cases, 
namely 1:6:1, 1:6:8, 1:12:16, is not yet optimal, 
in the sense that the expressions are quite intricate. 
The reason is that one could not
yet associate each of these three cases to an optimal PDE system
\cite{BakerThesis}. 
The link established in this paper between a PDE
with source system and one of the cubic H\'enon-Heiles Hamiltonians
strongly suggests that there could exist three privileged systems of
the type PDE+source, whose reduction $x-ct$ would be identical to
the Hamilton equations of the quartic cases 1:6:1, 1:6:8, 1:12:16. 
A preliminary step in this direction has been made in
\cite[Eq.~(A.24)]{CMBook} for the 1:6:1 and 1:6:8 cases, whose
integration with the autonomous F-VI equation \cite{CosPole2} needs
to be improved.

\section*{Acknowledgements }

The authors are grateful to the referees for quite valuable comments. 
Both authors would like to thank Prof.~Yunbo Zeng, Xingbiao Hu
and Runliang Lin for their good discussion and suggestions.
Meanwhile the author Junxiao Zhao wishes to express her gratitude
for the hospitality received when she was a postdoc at SPEC. 
This work is supported by the CAS President Grant, the National Natural
Science Foundation of China (Grant no.10701076).


\vfill \eject
\end{document}